\documentclass[preprint,12pt]{elsarticle}
\usepackage{ifpdf}
\usepackage{graphicx}
\usepackage{natbib}
\usepackage{amssymb}
\usepackage{amsmath}
\usepackage{epsfig}
\usepackage{graphics}
\journal{Communications in Nonlinear Science and Numerical Simulation}
\begin{document}

\begin{frontmatter} 

\title{Ghost-vibrational resonance}
\author{S~Rajamani}
\ead{rajeebard@gmail.com}
\author{S~Rajasekar\corref{auth2}}
\ead{rajasekar@cnld.bdu.ac.in}
\address{School of Physics, Bharathidasan University, 
Tiruchirappalli  620 024, India}
\author{MAF~Sanju\'{a}n}
\ead{miguel.sanjuan@urjc.es}
\address{Nonlinear Dynamics, Chaos and Complex Systems Group, Departamento de F\'{i}sica,
 Universidad Rey Juan Carlos, Tulip\'{a}n s/n, 28933 M\'{o}stoles, Madrid, Spain}





\begin{abstract}
Ghost-stochastic resonance is a noise-induced resonance at a fundamental frequency missing in the input signal. We investigate the effect of a high-frequency, instead of a noise, in a single Duffing oscillator driven by a multi-frequency signal  $F(t)= \sum^n_{i=1} f_i \cos(\omega_i + \Delta \omega_0)t$, $\omega_i=(k+i-1)\omega_0$, where $k$ is an integer greater than or equal to two. We show the occurrence of a high-frequency induced resonance at the missing fundamental frequency $\omega_0$. For the case of the two-frequency input signal, we obtain an analytical expression for the amplitude of the periodic component with the missing frequency. We present the influence of the number of forces $n$, the parameter $k$, the frequency $\omega_0$ and the frequency shift $\Delta \omega_0$ on the response amplitude at the frequency $\omega_0$. We also investigate the signal propagation in a network of unidirectionally coupled Duffing oscillators. Finally, we show the enhanced signal propagation in the coupled oscillators in absence of a high-frequency periodic force.
\end{abstract}


\end{frontmatter}
\section{Introduction}
Response of nonlinear systems to a harmonic force with a single frequency has been investigated in detail. A non-monotonic variation of the amplitude of the response occurs \cite{{r1},{r2}}, in a typical nonlinear system when the frequency of the driving force is varied. In particular, the oscillation amplitude of the system output increases with the increase in the frequency of the external force, it reaches a maximum at a particular frequency and then it decreases with further increase in the frequency. This resonance phenomenon is widespread and has been utilized in several devices. In bistable and multistable systems when the amplitude of the external periodic force is below a threshold (that is, there is no switching motion between the coexisting stable states), then a transition between the coexisting states can be induced by a weak noise. At an appropriate optimum noise intensity, almost a periodic switching between coexisting states occurs resulting in a maximum system response. This noise-induced resonance phenomenon is termed as {\emph{stochastic resonance}} \cite{{r3},{r4}}. Resonance can be realized when the noise term is replaced by a high-frequency periodic force and is called {\emph{vibrational resonance}} \cite{{r5},{r6}}. Furthermore, it is possible to generate a chaotic signal that mimics the probability distribution of the Gaussian white noise. Such a signal can also give rise to a resonant effect analogous to the noise-induced resonance and is called {\emph{chaotic resonance}} \cite{{r7}}.
In all the above resonance phenomena, in absence of a resonance inducing source, the system is driven by a weak harmonic force with a single frequency. There are signals with multiple frequencies. Examples include human speech, musical tones and square-waves. Design of an approximate multi-frequency signal is very important in minimizing the nonlinear distortion in the multi-frequency system identification methods \cite{{r8},{r9}}.

Chialvo et al. \cite{{r10},{r11}} investigated the response of a threshold device to an input signal containing several frequencies in the presence of noise. When the frequencies of the driving force are of a higher-order of a certain fundamental frequency, then the system is found to show a maximum response at the missing fundamental frequency at an optimum noise intensity. This fundamental frequency, which is absent in the input signal, detected by the device is called ghost-frequency and the underlying resonance phenomenon is termed as {\emph{ghost-stochastic resonance}} \cite{{r10},{r11}}. When the input signal is set into an anharmonic by introducing a same frequency shift to all the harmonic terms, the system is found to show a resonance at a certain shifted frequency. This ghost resonance phenomenon can be used to explain the missing fundamental illusion in which a third lower pitched tone is often heard when two tones occur together \cite{{r11}}.

The occurrence of a ghost resonance induced by noise has been analysed mostly in excitable systems. For example, it was found in the sudden dropouts exhibited by a semiconductor laser \cite{r12}, two laser systems coupled bidirectionally \cite{r13}, vertical-cavity surface emitting lasers \cite{r14}, monostable Schmitt trigger electronic circuit \cite{r15}, an excitable Chua's circuit \cite{r16}, a chaotic Chua's circuit \cite{r17} and a system of $n$-coupled neurons \cite{r18}. Subharmonic resonance behaviour in a nonlinear system with a multi-frequency force containing the fundamental frequency in the absence of a high-frequency input signal is studied in \cite{r19}.

Because nonlinear systems with double-well and multi-well potentials are wide-spread it is foremost important to investigate the response of these systems to the multi-frequency force and analyse the occurrence of ghost resonance in them and also with sources other than external noise. Motivated by the above considerations, in the present work, we explore the possibility of a ghost resonance induced by a high-frequency deterministic force rather than a noise. We consider the Duffing oscillator driven by multi-frequency force $F(t)$ and a high-frequency force $g\cos\Omega t$. The multi-frequency force $F(t)$ is given by
\begin{equation}
     F(t) = \sum^n_{i=1} f_i \cos(\omega_i
              + \Delta \omega_0)t, 
               \quad \omega_i=(k+i-1) \omega_0 \label{eq1}
\end{equation}
with $k \ge 2$ and $\Omega \gg \omega_n (=(k+n-1) \omega_0)$. We begin our analysis with $n=2$, $k=2$ and $\Delta \omega_0=0$. We show the occurrence of a resonance at the fundamental frequency $\omega_0$ missing in the input signal $F(t)$. The value of $g$ at which the resonance at the frequency $\omega_0$ occurs, increases monotonically while the value of the response amplitude $Q(\omega_0)$ at resonance decreases with $\omega_0$. Interestingly, the case of $n=2$ by applying a theoretical method, we are able to obtain an approximate analytical expression for the response amplitudes $Q(\omega_i)$, $i=0,1,2$. Theoretical results are in good agreement with the numerical predictions. We study the influence of the number of periodic forces $n$, the parameters $k$ and $g$ and the frequency shift $\Delta \omega_0$ on $Q(\omega_0)$. For values of $k>2$ or $\Delta \omega_0 \ne 0$, the response amplitude $Q(\omega_0)$ becomes $0$ when the oscillation center of the orbit is at the origin and this happens for $g$ values above a certain critical value.

Next, we consider a network of unidirectionally coupled $N$-Duffing oscillators with the multi-frequency force and the high-frequency force applied to the first oscillator only. The first system is uncoupled. The coupling term is chosen to be linear. We denote $Q_i(\omega_0)$ as the response amplitude of the $i$th oscillator at the frequency $\omega_0$. For a coupling strength above a critical value, an undamped signal propagation, that is, $Q_N(\omega_0) > Q_1(\omega_0)$ occurs at the missing fundamental frequency, even in the absence of the high-frequency periodic force. Interestingly, in the undamped signal propagation case, the response amplitude increases with the unit number $i$ and then becoming a constant. The saturation value of $Q$ is found to be independent of the parameters $k$, $n$ and $\Delta \omega_0$ in $F(t)$. Finally, we consider a network of unidirectionally coupled oscillators, where all the oscillators are driven by the external forces.
\section{Resonance in a single Duffing oscillator}
 \label{rsdo}
We consider the equation of motion of the Duffing oscillator driven by $n$ harmonic forces $F(t)$ given by Eq.~(\ref{eq1}) and the high-frequency periodic force $g \cos \Omega t$ as 
\begin{eqnarray}
\label{eq2}
    \ddot x + d \dot x + \alpha x + \beta x^3
       = F(t) + g \cos \Omega t .
\end{eqnarray}
Throughout our study we fix the values of the parameters as $d=0.5$, $\alpha=-2$, $\omega_0=0.5$, $\beta=1$, $\Omega =30 \omega_0$ and treat $g$ as the control parameter. The potential associated to the system in the absence of damping and external force is of a double-well form, since $\alpha < 0$ and $\beta > 0$.
\subsection{Numerical analysis}
From the numerical solution of Eq.~(\ref{eq2}), we compute the sine and cosine components $Q_{\mathrm{s}}(\omega)$ and $Q_{\mathrm{c}}(\omega)$ respectively of the solution at various frequencies in the interval $\omega \in [0,20]$ using the equations
\begin{subequations}
 \label{eq3}
\begin{eqnarray}
    Q_{\mathrm{s}}(\omega) 
       & = & \frac{2}{NT} \int^{NT}_0 x(t) \sin \omega t\, dt, \\
    Q_{\mathrm{c}}(\omega)
       & = & \frac{2}{NT} \int^{NT}_0 x(t) \cos\omega t\, dt,
\end{eqnarray}
\end{subequations}
where $T = 2 \pi / \omega$ and $N$ is say $500$. Then $Q(\omega) = \sqrt{Q^2_{\mathrm{s}} + Q^2_{\mathrm{c}}} / f$ with $f = (1/n)\sum^n_{i=1} f_i$.

\begin{figure}[t]
\begin{center}
\epsfig{figure=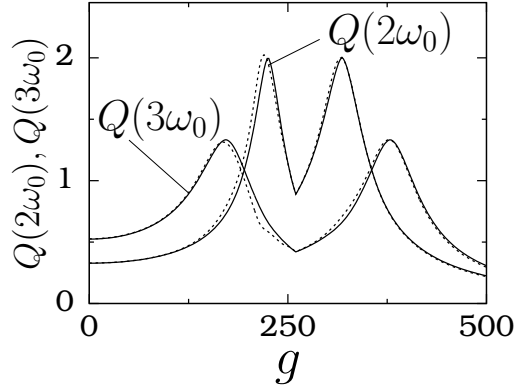, width=0.5\columnwidth}
\end{center}
\caption{$Q(2\omega_0)$ and $Q(3\omega_0)$ versus $g$ of the system (\ref{eq2}) with $F(t)=f_1 \cos 2\omega_0t + f_2 \cos 3\omega_0t$ for the cases (i) $f_1=0.1$, $f_2=0$ ($Q(2 \omega_0) \ne 0$ (continuous curve), $Q(3\omega_0)=0$), (ii) $f_1=0$, $f_2=0.1$ ($Q(2\omega_0)=0$, $Q(3\omega_{0}) \ne 0$ (continuous curve)) and (iii) $f_1=f_2=0.1$ (both $Q(2\omega_{0})$ and $Q(3\omega_{0})$ are nonzero and are represented by dashed curves).}
 \label{f1}
\end{figure}

First, we consider the system (\ref{eq2}) with $n=2$, $k=2$ and $\Delta \omega_0=0$, that is $F(t)=f_1 \cos \omega_1 t + f_2 \cos \omega_2 t$ where $\omega_1 = 2 \omega_0$, $\omega_2 = 3 \omega_0$ and $\omega_0=0.5$. We compute $Q(\omega_0)$, $Q(2\omega_0)$ and $Q(3\omega_0)$ for (i) $f_1=f$, $f_2=0$, (ii) $f_1=0$, $f_2=f$ and (iii) $f_1=f_2=f$ with $f=0.1$. Figure \ref{f1} shows the variation of numerically computed $Q(2\omega_0)$ with the parameter $g$ for the cases (i) and (iii) (represented by continuous and dashed curves respectively) and $Q(3\omega_0)$ for the cases (ii) and (iii) (continuous and dashed curves respectively). For $f_2=0$, $Q(3\omega_0)=0$ while for $f_1=0$, $Q(2\omega_0)=0$. In both cases $Q(\omega_0)=0$. When $f_1 \ne 0$ and $f_2 \ne 0$ both $Q(2\omega_0)$ and $Q(3\omega_0)$ are present in the solution of the system (\ref{eq2}). $Q$ at $\omega = 2\omega_0$ and $3\omega_0$ exhibits resonance. For a wide range of $g$, $Q(2\omega_0)$ of the cases (i) and (iii) are almost the same. This result is observed for $Q(3\omega_0)$ except for the values of $g$ near the first resonance of $Q(2\omega_0)$. We can say that there is no significant effect of the presence of the periodic force $f_1 \cos 2\omega_0 t$ on $Q(3\omega_0)$ and $f_2 \cos 3 \omega_0 t$ on $Q(2\omega_0)$. However, in the presence of these two periodic low-frequency forces and with $g=0$ the solution of the system (\ref{eq2}) contains periodic components with certain frequencies other than $2 \omega_0$ and $3 \omega_0$. However, $Q$ at these frequencies are very weak.

\begin{figure}[t]
\begin{center}
\epsfig{figure=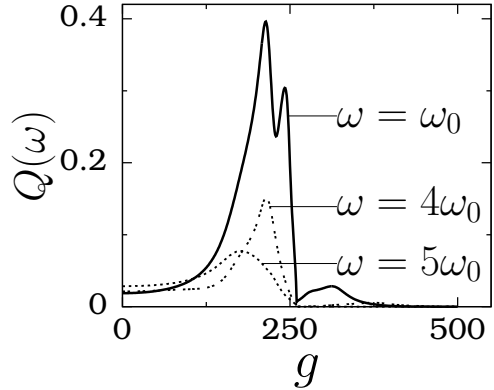, width=0.48\columnwidth}
\end{center}
\caption{Variation of $Q(\omega)$ with $g$ in the system (\ref{eq2}) for the frequencies $\omega = \omega_0$ $(=0.5)$, $4\omega_0$ and $5\omega_0$ missing in the input signal $F(t)$ (Eq.~(\ref{eq1})) with $n=2$, $k=2$, $\Delta \omega_0=0$ and $f_1=f_2=f=0.1$.}
 \label{f2}
 \end{figure}

Figure \ref{f2} presents $Q(\omega)$ versus $g$ for $\omega = \omega_0$, $4\omega_0$ and $5\omega_0$. In this figure, we find that $Q(\omega_0) \ne 0$ in the absence of a high-frequency force ($g=0$). However, its value is $\approx 0$. When $g$ is varied $Q(\omega)$ at $\omega = \omega_0$, $4\omega_0$, $5\omega_0$ exhibits a resonance. The resonance of $Q(\omega_0)$ is relatively stronger than at the frequencies $4\omega_0$ and $5\omega_0$. We note that the fundamental frequency $\omega_0$ is missing in the input signal $F(t)$. The resonance phenomenon induced by an external noise at a frequency that is absent in the input signal is termed as {\emph{ghost-stochastic resonance}} \cite{{r10},{r11}}. We call the high-frequency deterministic force induced resonance at the missing frequency of the input signal as {\emph{ghost-vibrational resonance}}. There are two fundamental differences between the ghost resonance induced by noise and by the high-frequency force. In the noise driven case, when the intensity $D$ of the noise is varied, the signal-to-noise ratio at a missing fundamental frequency becomes maximum at one value of $D$. Further, the resonances at the frequencies present in the input signal are weak. In the high-frequency induced ghost resonance, the response amplitude can be maximum at more than one value of the parameter $g$ (as shown in Fig.~\ref{f2}). In the system (\ref{eq2}), the resonance at the frequencies present in the input signal are stronger than the resonance at the missing fundamental frequency. The resonance at the frequencies $2\omega_0$ and $3\omega_0$ is the well known vibrational resonance.

\begin{figure}[t]
\begin{center}
\epsfig{figure=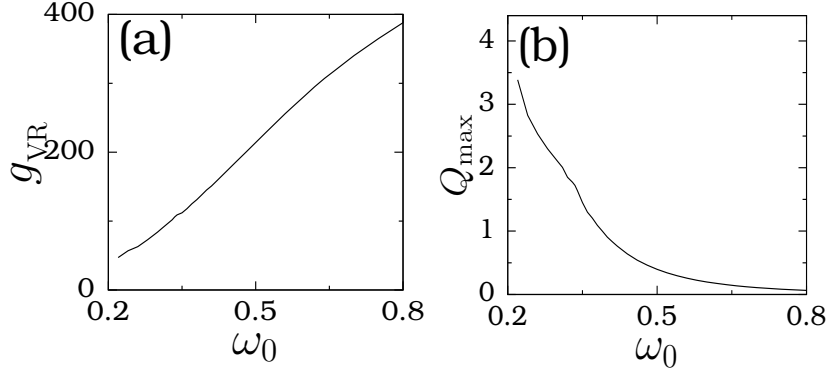, width=0.8\columnwidth}
\end{center}
\caption{Variation of (a) $g_{_{\mathrm{VR}}}$, the value of $g$ at which first resonance at the frequency $\omega_0$ occurs, and (b) $Q_{\mathrm{max}}(\omega_0)=Q(g_{_{\mathrm{VR}}}, \omega_0)$ with the missing fundamental frequency $\omega_0$ in the input signal in the system (\ref{eq2}). The values of the parameters in (\ref{eq1}) are as in Fig.~\ref{f2}.}
\label{f3}
\end{figure}

We numerically compute $g_{_{\mathrm{VR}}}$, the value of $g$ at which a first resonance occurs and the corresponding value of the response amplitude, $Q_{\mathrm{max}}$, for a range of values of $\omega_0$. The result is shown in Fig.~\ref{f3}. $g_{_{\mathrm{VR}}}$ increases almost linearly with $\omega_0$ while $Q_{\mathrm{max}}$ decreases nonlinearly with $\omega_0$.

%
\subsection{Theoretical calculation of $Q(\omega_0)$}
It is possible to obtain analytical expressions for the response amplitudes $Q(\omega)$ at various values of $\omega$. For the system (\ref{eq2}) with $n=2$ we assume that its solution consists of a low-frequency component $X$ and a high-frequency $(\Omega)$ component $\psi$. Substituting $x = X + \psi$ in Eq.~(\ref{eq2}) we obtain
\begin{eqnarray}
 \label{eq4}
    \ddot X + d \dot X + \alpha X + \beta X^3
       + 3 \beta X^2 \langle \psi \rangle 
       + 3 \beta X \langle \psi^2 \rangle 
     =  f_1 \cos \omega_1 t + f_2 \cos\omega_2 t,
\end{eqnarray}
\begin{eqnarray}
 \label{eq5}
    \ddot \psi + d \dot \psi + \alpha \psi + \beta \psi^3
      + 3\beta X^2 \left( \psi - \langle \psi \rangle \right)
      + 3 \beta X \left( \psi^2 - \langle \psi^2 \rangle \right)
      = g \cos \Omega t ,
\end{eqnarray}
where $\langle \psi^m \rangle = (1 / 2 \pi) \int_0^{2\pi} \psi^m d\tau$ and $\tau = \Omega t$.  Since $\psi$ is rapidly oscillating, it is reasonable to approximate the Eq.~(\ref{eq5}) as $\ddot{\psi} = g \cos \Omega t$, which gives $\psi = - (g / \Omega^2) \cos \Omega t$. For this solution $\langle \psi \rangle = 0$, $\langle \psi^2 \rangle = g^2 / (2\Omega^4)$ and $\langle \psi^3 \rangle =0$. Then, Eq.~(\ref{eq4}) becomes
\begin{eqnarray}
\label{eq6}
   \ddot X + d \dot X + C X + \beta X^3
     =f_1 \cos \omega_1 t + f_2 \cos \omega_2 t,
\end{eqnarray}
where $C = \alpha + 3 \beta g^2 / (2\Omega^4)$. Slow oscillations of (\ref{eq6}) occur about its stable equilibrium points. Equation (\ref{eq6}) with $f_1=f_2=0$ admits three equilibrium points
\begin{eqnarray}
 \label{eq7}
     X^* = 0, ~\pm \sqrt{-C/\beta} ~~ \text{for} ~~ 
        g < g_{\mathrm{c}} = \left[
         - \frac{2 \alpha \Omega^4}{3\beta} \right]^{1/2}
\end{eqnarray}
and only one equilibrium point $X^*=0$ for $g > g_{\mathrm{c}}$. For convenience, we introduce the change of variable $Y=X-X^*$. This gives
\begin{subequations}
 \label{eq8}
\begin{eqnarray}
  \ddot{Y} + d \dot{Y} + \omega^2_{\mathrm{r}} Y 
     + \beta Y^3 + 3 \beta Y^2 X^*
     = f_1 \cos \omega_1 t + f_2 \cos \omega_2 t,
\end{eqnarray}
where
\begin{eqnarray}
    \omega^2_{\mathrm{r}} = \alpha
       + \frac{3\beta g^2}{2\Omega^4} + 3\beta X^{*2}.
\end{eqnarray}
\end{subequations}

For a weak nonlinearity, an approximate solution of Eq.~(\ref{eq8}) can be constructed through an iterative process \cite{r2}, wherein we obtain the sequence of approximations $Y_0(t)$, $Y_1(t)$,$\cdots$ by solving the equations
\begin{eqnarray}
 \label{eq9}
     \ddot{Y}_0 + d \dot{Y}_0 + \omega^2_{\mathrm{r}} Y_0
         & = & F(t), \\
     \label{eq10}
     \ddot{Y}_1 + d \dot{Y}_1 + \omega^2_{\mathrm{r}} Y_1
         & = & F(t) - \beta Y_0^3 - 3 \beta X^* Y^2_0 
\end{eqnarray}
and so on. We determine both $Y_0$ and $Y_1$. The solution of Eq.~(\ref{eq9}) in the long time limit is 
\begin{subequations}
 \label{eq11}
\begin{eqnarray}
    Y_0(t) = A_1 \cos(\omega_1 t + \phi_1)
              + A_2 \cos ( \omega_2 t + \phi_2 ),
\end{eqnarray}
where
\begin{eqnarray}
    A_i  =  \frac{f}{\sqrt{\left(\omega^2_{\mathrm{r}}
                -\omega^2_i \right)^2 
                 + d^2 \omega^2_i}}, \quad
   \phi_i  =  \tan^{-1} 
                \left( - \frac{d\omega_i}{\omega^2_{\mathrm{r}}
                   -\omega^2_i } \right), \;\;i=1,2.
\end{eqnarray}
\end{subequations}
Substituting the above expression for $Y_0$ in (\ref{eq10}), we can find the solution $Y_1$. In addition to the frequencies $\omega_1$ and $\omega_2$, the solution $Y_1$ contains certain other frequencies, namely, $l \omega_0$, where $l=1$, $k-1$, $k$, $k+1$, $k+2$, $2k$, $2k+1$, $2k+2$, $3k$, $3k+1$, $3k+2$, $3k+3$ due to the terms $Y^2_0$ and $Y^3_0$ in Eq.~(\ref{eq10}). When $k=2$ the various frequencies present in $Y_1$ are $\omega=l\omega_0$, $l=1,2,\cdots,9$. The lowest and the highest frequencies in $Y_1$ are $\omega_2 - \omega_1 = \omega_0$ and $3\omega_2$ respectively.

Retaining only the terms containing $\omega_0$, $\omega_1$, $\omega_2$, $\omega_1 - \omega_0$ (which will become $\omega_0$ if $k=2$) in the right-side of Eq.~(\ref{eq10}) we obtain (with $Y_1$= $Y_1$ ($\omega_0$, $\omega_1$, $\omega_2$, $\omega_1 - \omega_0$))
\begin{subequations}
 \label{eq12}
\begin{eqnarray}
   Y_1 & = &  \frac{a_{01}}{s_0} 
              \cos \left(\omega_0t + \phi_0 - \phi_1
              + \phi_2 \right)   \nonumber \\
       &   &  + \frac{a_{02}}{s_{01}} \cos \left( 
              ( \omega_1 - \omega_0 ) t + \phi_0 
              + 2 \phi_1 - \phi_2 \right) \nonumber \\
       &   &  + \frac{a_1}{s_1} \cos \left( 
              \omega_1 t + \phi_1 + \phi_2 \right)
              + \frac{f}{s_1} \cos \left( \omega_1 t
              + \phi_1 \right)  \nonumber \\
       &   &  + \frac{a_2}{s_2} \cos \left( 
              \omega_2 t + 2 \phi_2 \right) 
              + \frac{f}{s_2} \cos \left( \omega_2 t
              + \phi_2 \right),
\end{eqnarray}
where
\begin{eqnarray}
    \phi_i  
    & = &  \tan^{-1} \left( 
           - \frac{d \omega_i}{\omega^2_{\mathrm{r}}
           - \omega^2_i} \right), \;\;\; i=0,1,2 \\
    \omega_i 
    & = &  ( k + i - 1 ) \omega_0 + \Delta \omega_0,
            \;\;\; i=1,2 \\
    s_i 
    & = &  \sqrt{ \left( \omega^2_{\mathrm{r}} 
           - \omega^2_i \right)^2 + d^2 \omega^2_i}, 
            \;\;\; i=0,1,2 \\
    s_{01} 
    & = &  s_0 \left( \omega_0 \to \omega_1 - \omega_0 \right),
           \;\; A_i =\frac{f}{s_i}, \;\;\; i=1,2 \\
    a_{01} 
    & = &  3 \beta X^* A_1 A_2, \;\; 
           a_{02} = \frac{3}{4} \beta A_1^2 A_2, \\
    a_1 
    & = &  \frac{3}{4} \beta A_1 \left( A_1^2 + 2A_2^2 \right),
           \;\; a_2 = \frac{3}{4} \beta A_ 2 
           \left( A_2^2 + 2 A_1^2 \right) .
\end{eqnarray}
\end{subequations}
For $k=2$ and $\Delta \omega_0=0$ we notice that $\omega_1 - \omega_0 = \omega_0$.  In this case  the first two terms in the right-side of Eq.~(\ref{eq12}a) are periodic with frequency $\omega_0$, otherwise the first term alone is periodic with a frequency $\omega_0$.

In Eq.~(\ref{eq12}a) for $\lvert f_1 = f_2 = f \lvert \ll 1 $ we can drop the third and fifth terms in the right-side because $a_1$ and $a_2$ are of the order of $f^3$, while the fourth and sixth terms are of the order of $f$ only and the minimum value of $s_i$ is $d \omega_i$ which is not very small for $d=0.5$ and $\omega_0=0.5$. Then the amplitude of the periodic components in the solution (\ref{eq12}a) with the frequencies $\omega_0$, $\omega_1$ and $\omega_2$ are 
\begin{subequations}
\label{eq13}
\begin{eqnarray}
   A (\omega_1) 
     & = &  \frac{f}{s_1}, \quad 
            A (\omega_2) = \frac{f}{s_2},  \\
   A ( \omega_0, k=2,\Delta \omega_0 = 0 ) 
     & = & \frac{\sqrt{a^2_{01} + a^2_{02}
            + 2 a_{01} a_{02} \cos (2 \phi_2 - 3 \phi_1)}}{s_0}\;, \nonumber  \\
              & &   \\
   A ( \omega_0, k \ne  2~{\mathrm{or}}~ 
      \Delta \omega_0 \ne 0) 
     & = &  \frac{a_{01}}{s_0} 
            = \frac{3 \beta X^* f^2}{s_0 s_1 s_2} .
\end{eqnarray}
\end{subequations}
Then $Q(\omega_i)=A(\omega_i)/f$.

\begin{figure}
\begin{center}
\epsfig{figure=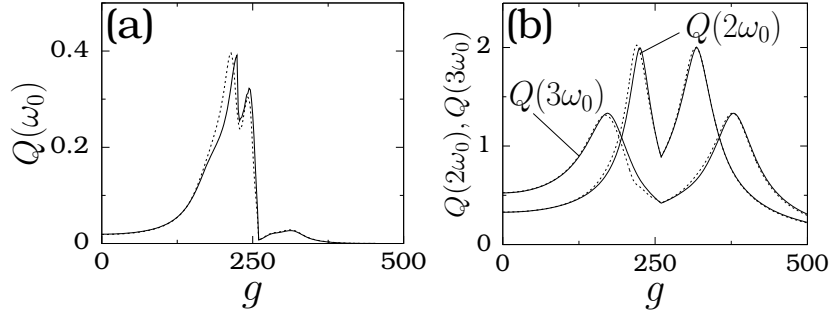, width=0.8\columnwidth}
\end{center}
\caption{Theoretically calculated (continuous curve) and numerically computed (dashed curve) $Q(\omega_0)$, $Q(2\omega_0)$ and $Q(3\omega_0)$ versus $g$ for the system (\ref{eq2}) with $\omega_0=0.5$, $n=2$, $k=2$, $\Delta \omega_0=0$ and $f_1=f_2=f=0.1$. }
\label{f4}
\end{figure}

To verify the theoretical treatment, we plot in Fig.~\ref{f4} both theoretically and numerically calculated $Q(\omega_0)$, $Q(2\omega_0)$ and $Q(3\omega_0)$ as a function of the parameter $g$. We notice that theoretical $Q$ at $\omega=\omega_0$, $2\omega_0$ and $3\omega_0$ are in very good agreement with the numerically computed $Q$. 

$Q(\omega_i)$ becomes maximum when $\omega^2_{\mathrm{r}} = \omega^2_i$, $i=1,2$. Then, the analytical expressions for $g$ at which resonances occur, denoted as $g_{_{\mathrm{VR}}}$, are given by
\begin{subequations}
\label{eq14}
\begin{eqnarray}
   g_{_{\mathrm{VR}}}^{(1)} (\omega_i) 
     & = & \Omega^2 \left[ \frac{1}{3\beta} 
            ( 2 \lvert \alpha \rvert - \omega^2_i)
            \right]^{1/2}, \;\; i=1,2 \\
   g_{_{\mathrm{VR}}}^{(2)} (\omega_i) 
     & = & \Omega^2 \left[ \frac{2}{3\beta} (\omega^2_i 
            + \lvert \alpha \rvert ) \right]^{1/2}, \;\; i=1,2
\end{eqnarray}
\end{subequations}
where $2 \lvert \alpha \rvert > \omega^2_i$. Because $A(\omega_0)$ is a complicated function of $g$ it is very difficult to find an analytical expression for $g_{_{\mathrm{VR}}}$ at $\omega_0$.
\section{Effect of $k$, $n$ and $\Delta \omega_0$ on resonance in the single Duffing oscillator}
 \label{ersdokn}
The theoretical procedure employed in the previous section for the determination of an analytical expression for $Q(\omega_0)$ can be extended for $n > 2$. Since such an analysis involves tedious mathematics we perform a numerical simulation. We choose $f_i=f$, $i=1,2,\cdots,n$.

\begin{figure}
\begin{center}
\epsfig{figure=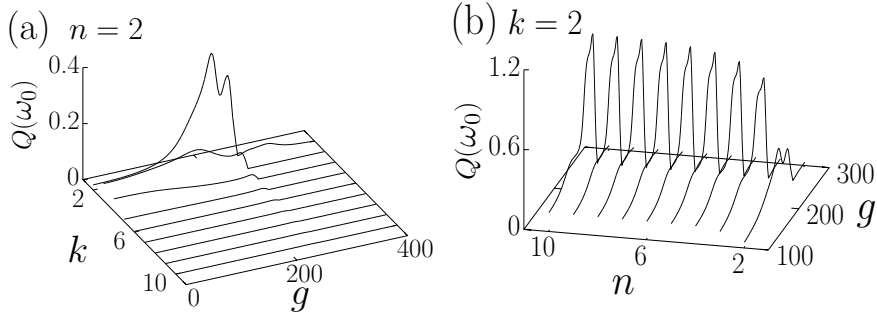, width=0.85\columnwidth}
\end{center}
\caption{Three-dimensional plot of $Q(\omega_0)$ versus $g$ and (a) $k$ for $n=2$ and (b) $n$ for $k=2$ for the system (\ref{eq2}) with $\omega_0=0.5$ and $f=0.1$.}
\label{f5}
\end{figure}

Figure \ref{f5}(a) presents $Q(\omega_0)$ versus $g$ for several values of $k$ with $n=2$ and $\Delta \omega_0=0$. $Q(\omega_0)$ (as well as $Q(\omega_1)$ and $Q(\omega_2)$) decays to zero with $k$. For $k \ne 2$ the theoretical expression for $A(\omega_0)$ in $Q(\omega_0)=A(\omega_0)/f$ is given by Eq.~(\ref{eq13}c). For a fixed value of $g$ as $k$ increases the quantities $s_i$, $i=0,1,2$ increase and $A_1$ and $A_2$ decrease. Since $A(\omega_0)$ is directly proportional to $A_1$, $A_2$ and $1 / s_0$, its value decreases with increasing values of $k$. We notice in Fig.~\ref{f5}(a)  that $Q(\omega_0)=0$ for $g > g_{\mathrm{c}}$($=259.81$) when $k>2$. This is because for $g>g_{\mathrm{c}}$ the equilibrium point about which a slow oscillation takes place is $X^*=0$ and hence $Q(\omega_0)$ becomes zero (refer Eq.~(\ref{eq13}c)). That is, for $k>2$ the output signal will have a periodic component with the missing frequency $\omega_0$ only if the center of oscillation of the output $x(t)$ is $\ne 0$ which will happen for $g < g_{\mathrm{c}}$. We note that $g_{\mathrm{c}}$ (given by Eq.~(\ref{eq7})) depends on the parameters $\alpha$, $\beta$ and $\Omega$.
 
The value of $Q(\omega_0)$ at resonance, as shown in Fig.~\ref{f5}(b), increases with the number of periodic forces, $n$, and attains a saturation. For $k>2$ resonance occurs for $n \ge 2$, but the value of $Q(\omega_0)$ at resonance decreases when $n$ increases.

\begin{figure}[t]
\begin{center}
\epsfig{figure=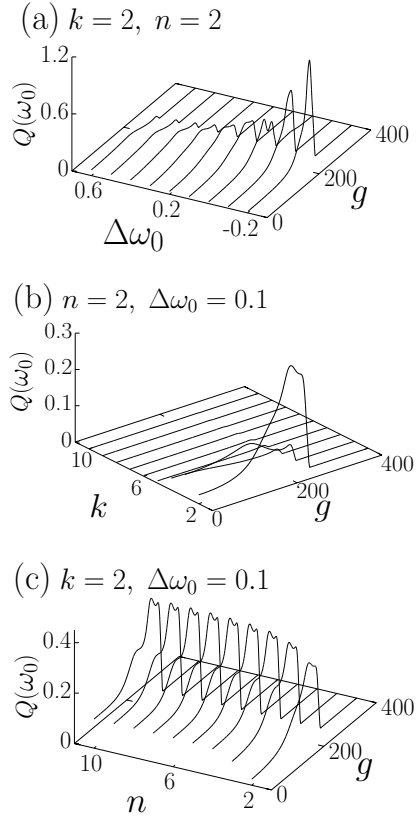, width=0.4\columnwidth}
\end{center}
\caption{$Q(\omega_0)$ versus $g$ for various values of (a) $\Delta \omega_0$ for $k=2$, $n=2$, (b) $k$ for $n=2$, $\Delta \omega_0=0.1$ and (c) $n$ for $k=2$, $\Delta \omega_0=0.1$. In all the cases $\omega_0 = 0.5$ and $f=0.1$.}
\label{f6}
\end{figure}

Next, we consider the system (\ref{eq2}) with $\Delta \omega_0 \neq 0$. When $\Delta \omega_0=0$ the frequencies of the periodic forces in $F(t)$ are integer multiples of the fundamental frequency $\omega_{0}$. For $\Delta \omega_0\ne 0$ the frequency difference between successive periodic components remains the same. The frequencies of the periodic components are essentially shifted multiples of $\omega_0$. Each component in $F(t)$ is periodic while the force $F(t)$ is aperiodic, that is, anharmonic. Figure \ref{f6}(a) displays the effect of $\Delta \omega_0$ on the response the amplitude profile for $k=2$, $n=2$ and $\omega_0 = 0.5$. For $\Delta \omega_0 \ne 0$ the amplitude $A(\omega_0)$ given by Eq.~(\ref{eq13}c) is inversely proportional to $s_0 s_1 s_2$. For a fixed value of $g$, the quantities $s_i$, $i=0,1,2$ increase with increase in $\Delta \omega_0$. Consequently, $A(\omega_0)$ and hence $Q(\omega_0) = A(\omega_0) / f$ decrease with increase in $\Delta \omega_0$. This is evident in Fig.~\ref{f6}(a). We observe a similar result in Fig.~\ref{f6}(b) where $Q(\omega_0)$ is plotted as a function of $k$ and $g$ for $n=2$ and $\Delta \omega_0=0.1$. In Fig.~\ref{f6}(c) $Q(\omega_0)$ versus $g$ for various values of $n$ is plotted for $k=2$ and $\Delta \omega_0=0.1$ where $Q(\omega_0)$ attains a saturation. This result is similar to the one shown in Fig.~\ref{f5}(b) for $\Delta \omega_0=0$. In all the subplots in Fig.~\ref{f6} $Q(\omega_0)=0$ for $g > g_{\mathrm{c}}$($=259.81$) because $A(\omega_0) \propto X^*$ and $X^*=0$ for $g > g_{\mathrm{c}}$.
%

\section{Signal propagation in one-way coupled systems}
 \label{spowcs}
In this section we analyse the features of signal propagation at the missing fundamental frequency in a regular network of one-way coupled $N(=200)$ units. We consider the cases of multi-frequency signal applied to (i) first unit only and (ii) to all the units.
\subsection{Description of the network model}
The network essentially consists of $N$ units. The first unit is uncoupled and is alone driven by both a multi-frequency input periodic signal and the high-frequency periodic signal. The interaction is along one direction. We choose the coupling term to be linear and the system representing each unit as the Duffing oscillator. The equation of motion of the network is given by
\begin{subequations}
 \label{eq15}
\begin{eqnarray}
   \ddot{x}_1 + d \dot{x}_1 + \alpha x_1 + \beta x^3_1
      & = &  F(t) + g \cos \Omega t,  \\
   \ddot{x}_i + d \dot{x}_i + \alpha x_i + \beta x^3_i
      & = &  \delta x_{i-1}, \;\;\; i=2,3,\cdots,N
\end{eqnarray}
\end{subequations}
and $F(t)$ is given by Eq.~(\ref{eq1}). The dynamics of the first oscillator is independent of the dynamics of the other oscillators. We fix the values of the parameters in the network as $d=0.5$, $\alpha=-2$, $\beta=1$, $\omega_0=0.5$ and $\Omega = 30 \omega_0$. 
\subsection{Undamped signal propagation}
\begin{figure}[t]
\begin{center}
\epsfig{figure=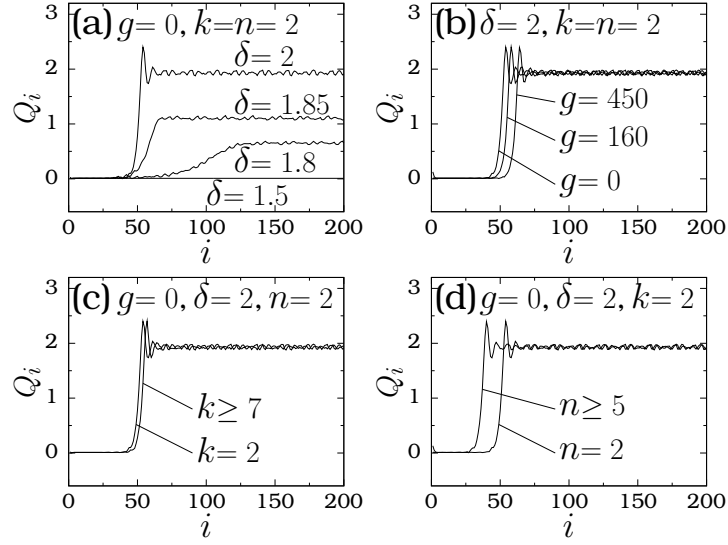, width=0.7\columnwidth,clip}
\end{center}
\caption{Dependence of $Q_i$ versus $i$ curve on (a) the coupling strength $\delta$ for $g=0$, $k=n=2$, (b) the amplitude $g$ of the high-frequency periodic force for $\delta=2$, $k=n=2$, (c) the parameter $k$ in $F(t)$ for $\delta=2$, $n=2$ and $g=0$ and (d) the number of periodic forces in $F(t)$ for $\delta=2$, $g=0$ and $k=2$. In all the cases $\omega_0=0.5$, $f=0.1$, $\Delta \omega_0=0$ and $\Omega=30\omega_0$. The first oscillator alone is driven by the force $F(t)$ and $g \cos \Omega t$. }
\label{f7}
\end{figure}

In the input signal $F(t)$ the fundamental frequency $\omega_0$ is absent. We numerically calculate the response amplitude $Q_i(\omega_0)$ using Eq.~(\ref{eq3}). Figure \ref{f7}(a) shows $Q_i(\omega_0)$ as a function of the unit $i$ for a few fixed values of the coupling strength $\delta$ with the values of the parameters in $F(t)$ as $n=2$, $k=2$, $\omega_0=0.5$, $\Delta \omega_0=0$,  $f_1 = f_2 = \cdots = f_n = f = 0.1$ and $g=0$. 
In the absence of the high-frequency force $Q_1 = 0.01874$. For $\delta < \delta_{\mathrm{c}} = 1.78$, $Q_i < Q_1$ for $i \gg 1$. For $\delta \ge \delta_{\mathrm{c}}$ as $i$ increases the value of $Q_i$ increases slowly then increases rapidly and reaches a saturation. $Q_{200} > Q_1$ and the network displays undamped signal propagation. For very large $i$ the response amplitude $Q_i$ oscillates about a value with small amplitude. Neglecting this small oscillation in $Q_i$, we notice that $Q_i$ becomes almost constant for sufficiently large values of $i$. We denote $Q_{\mathrm{L}}$ as the limiting (saturation) value of $Q_i$. In Fig.~\ref{f7}(a) $Q_{\mathrm{L}}$ increases with increase in $\delta$ from $\delta_{\mathrm{c}}$. The undamped and enhanced propagation of signal with the frequency $\omega_0$ missing in the input signal takes place even in the absence of the high-frequency periodic force. The enhanced signal propagation is due to the unidirectional coupling. Note that the input signal $F(t)$ is applied to the first oscillator only.

We plot $Q_i$ versus $i$ for three values of $g$ with $\delta=2$ in Fig.~\ref{f7}(b). $Q_{\mathrm{L}}$ is independent of the amplitude $g$. The values of $g$ have a strong influence on $Q_i$ only over a certain range of values of $i$ denoting the oscillators number. In Fig.~\ref{f7}(b) roughly in the $40$th to $70$th oscillators $Q_i$ varies with $g$. In this interval of $i$, $Q_i$ rapidly increases with $i$. As $g$ increases from $0$ the $Q_i$ profile oscillates and becomes stationary for sufficiently large values of $g$. Similar effects are found for various fixed values of the parameter $k$ and the number of periodic forces $n$. Figures \ref{f7}(c) and (d) report the influence of $k$ and $n$ respectively on $Q_i$ for $\delta=2$ and $g=0$. In Fig.~\ref{f7}(c) the $Q_i$ versus $i$ profile evolves to a stationary one with increase in the value of $k$. The $Q_i$ profile remains the same for $n \ge 5$ in Fig.~\ref{f7}(d). An interesting result is that $Q_{\mathrm{L}}$ is independent of $g$, $k$ and $n$ and depends on $\delta$. A numerical simulation is performed for $\Delta \omega_0 \ne 0$. Results similar to the case $\Delta \omega_0=0$ are observed. Furthermore, $Q_{\mathrm{L}}$ is found to be independent of $\Delta \omega_0$.
\subsection{A network with all the units driven by external forces}
\begin{figure}[t]
\begin{center}
\epsfig{figure=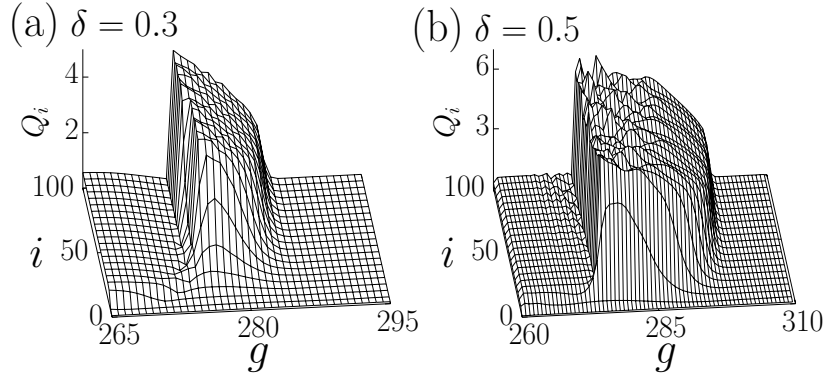, width=0.81\columnwidth}
\end{center}
\caption{Variation of $Q_i$ with $g$ and $i$ for two values of $\delta$ of the network system (\ref{eq16}) where $d=0.5$, $\alpha=-2$, $\beta=1$, $f=0.1$, $\omega_0 = 0.5$, $\Omega = 30\omega_0$, $k=2$, $n=2$ and $\Delta \omega_0 = 0$.}
\label{f8}
\end{figure}

\begin{figure}[!h]
\begin{center}
\epsfig{figure=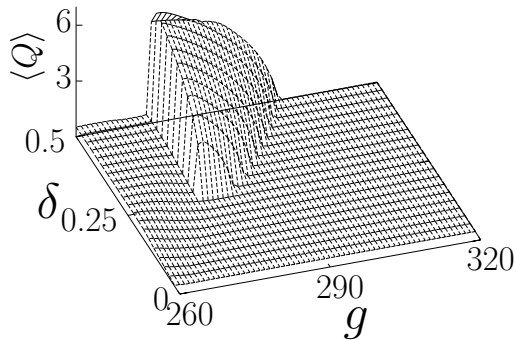, width=0.5\columnwidth}
\end{center}
\caption{Average response amplitude versus the control parameters $\delta$ and $g$ of the network (\ref{eq16}) with $f=0.1$, $\omega_0=0.5$, $\Omega=30\omega_0$, $k=2$, $n=2$ and $\Delta \omega_0=0$. }
\label{f9}
\end{figure}

Next, we consider the network with all the units driven by the force $F(t)$ and $g \cos \Omega t$ and the units are coupled unidirectionally. The equation of motion of the network is
\begin{subequations}
  \label{eq16}
\begin{eqnarray}
  \ddot{x}_1 + d \dot{x}_1 + \alpha x_1 
      + \beta x^3_1   & = & F(t) + g \cos \Omega t,  \\
  \ddot{x}_i + d \dot{x}_i + \alpha x_i 
      + \beta x^3_i & = & \delta x_{i-1} 
      + F(t) + g \cos \Omega t, \;\;\; i=2,3,\cdots, N.
\end{eqnarray}
\end{subequations}
Figure \ref{f8} shows $Q_i$ versus $g$ and $i$ for two fixed values of $\delta$. We observe ghost resonance in each unit. For $\delta=0.3$ (Fig.~\ref{f8}(a)) the value of $Q_i$ at resonance increases with the unit number $i$ and then reaches a saturation with $Q_{N,{\mathrm{max}}} > Q_{1,{\mathrm{max}}}$. For $\delta=0.5$ (Fig.~\ref{f8}(b)) we can clearly notice small oscillatory variation of $Q_i$ with $i$ for the values of $g$ near resonance. The oscillatory variation of $Q_i$ is found in the network system (\ref{eq15}) also. Comparing the Figs.~\ref{f7} and \ref{f8}, we observe that the enhancement of the response amplitude at resonance in the network (\ref{eq16}) is relatively higher than that of the network (\ref{eq15}). Figure~\ref{f9} presents the variation of $\langle Q \rangle = (1/N) \sum_{i=1}^N Q_i(\omega_0)$ with the parameters $\delta$ and $g$.


For the network system (\ref{eq15}), in Fig.~\ref{f7}(a) $Q$ of the last unit is $ < 3$ for $\delta < 2$. Furthermore, $Q$ of the last unit is independent of the value of $g$. For the network system (\ref{eq16}), in Fig.~\ref{f8} $Q$ of the last unit depends on the value of $g$. Further, in Fig.~\ref{f9} $\langle Q\rangle$ is $ > 3$ for a wide range of values of $\delta$. In the system (\ref{eq15}) undamped signal propagation with $Q_N > Q_1$ occurs for $\delta > 1.78$ and even for $g=0$. In contrast to this, in the system (\ref{eq16}) $Q_N > Q_1$ takes place only for certain range of values of $g$, however, $\langle  Q \rangle > Q_1$ even for a wide range of values of $\delta < 1.78$. As far as the signal amplification and propagation at a missing frequency $\omega_0$ is concerned, driving all units in the unidirectionally coupled system considerably improves the response amplitude over a certain range of values of $g$ and $\delta$ compared to the driving the first unit alone.

\section{Conclusion}
In a linear system driven by a single periodic force the output contains only the frequency present in the driving force. The response of a nonlinear system to a sinusoidal signal with a single frequency contains the input frequency and its harmonics. When a linear system is subjected to a multi-frequency force, the frequencies present in the output are the same as those in the input. However, changes occur in the magnitudes and phases of the various frequency components. In the case of a nonlinear system driven by a multi-frequency signal, the response not only contains the harmonics of the various input frequencies but inter-modulation components of harmonics can also be generated.

In the present work we have shown the enhancement of response amplitude of a nonlinear system at the missing fundamental frequency in the input multi-frequency signal. In the nonlinear system driven by multi-frequency force and noise, resonance at the missing fundamental frequency is the dominant one and it occurs at a relatively lower value of the noise intensity compared to the resonance at the frequencies present in the input signal. For these types of resonance to occur, the system must have a bistability or excitability. High-frequency induced ghost resonance can occur even in single-well nonlinear systems.

As shown in Fig.~\ref{f1}, the difference between, for example, $Q(2\omega_0)$ when the input signal contains only the frequency $2\omega_0$ and its value when other frequencies are also present in the input signal is negligible. That is, the response amplitude at a frequency $\omega$ present in the input signal is not affected appreciably by the presence of the other frequencies $\Omega_i$, as long as $\Omega_i$ are not widely separated from $\omega$. If any of the $\Omega_i \gg \omega$, then the vibrational resonance at the frequency $\omega$ occurs. 

When the input signal contains only a very few number of periodic components, then it is easy to obtain an analytical expression for the amplitudes of the periodic components with various frequencies. For the single oscillator, $Q(\omega_0)$ decays with the parameters $k$ and $\Delta \omega_0$, while it reaches a saturation with the number of forces. In the network where the oscillators are coupled unidirectionally and only the first oscillator is driven by the external forces, an enhanced and undamped signal propagation at the missing fundamental frequency takes place above a certain critical value of the coupling strength even in the absence of high-frequency force. Moreover, $Q_i(\omega_0)$ becomes constant for sufficiently large values of $i$. Finally, an interesting result is that, the limiting value of $Q$ is independent of the values of $g$, $k$ and $\Delta \omega_0$. 

 \vskip 5pt
\noindent{\bf{Acknowledgment}}
 \vskip 5pt
S.~Rajamani expresses her gratitude to University Grants Commission (U.G.C.), Government of India for financial support in the form of U.G.C. meritorious fellowship. MAFS acknowledges financial support from the Spanish Ministry of Science and Innovation under Project No. FIS2009-09898. 
\end{document}